\documentclass[12pt]{spieman}  
\usepackage{amsmath,amsfonts,amssymb}
\usepackage{graphicx}
\usepackage{setspace}
\usepackage{tocloft}

\title{The impact of station far sidelobes on EoR/CD power spectra}

\author[a,b]{Cathryn~M.~Trott}
\affil[a]{International Centre for Radio Astronomy Research, Curtin University, Bentley WA 6102, Australia}
\affil[b]{ARC Centre of Excellence for All Sky Astrophysics in 3D, Bentley WA 6102, Australia}

\cftpagenumbersoff{figure}
\cftpagenumbersoff{table} 
\begin{document} 
\maketitle

\begin{abstract}
Stations of dipole antennas for SKA1-Low will comprise 256 elements spread over an area with a diameter of 38~m. We consider the effect of residual unsubtracted sources well outside of the main beam for differing numbers of unique station configurations, in the Epoch of Reionisation (EoR) and the Cosmic Dawn (CD), both in simulation and with theoretical considerations. We find that beam sidelobes imprint power that renders the cosmological signal unobservable over a range of scales unless compact sources are subtracted beyond $\theta_Z=30$ degrees from zenith, and that station apodization will likely be required to control far sidelobes. An array with $N_b=4$ unique station configurations is sufficient to reduce the contamination, with an increase to $N_b=8$ showing little improvement. Comparison with an achromatic Airy disk beam model shows that beam sidelobe level is the main contributor to excess power in the EoR window, and beam chromaticity is less relevant. In the EoR, $z=8.5$, subtracting sources above 200~mJy out to $\theta_Z=45$ degrees, will be required to access relevant modes of the power spectrum.
\end{abstract}


{\noindent \footnotesize\textbf{*}Cathryn Trott,  \linkable{cathryn.trott@curtin.edu.au} }


\section{Introduction}
\label{sect:intro}  
Exploration of the first billion years of the Universe's history through observation of the neutral hydrogen hyperfine transition remains one of the highest priority science cases for the low-frequency array of the Square Kilometre Array (SKA1-Low, \citenum{koopmans15}). While the current generation of telescopes hopes to detect signals from the Cosmic Dawn (CD; 25$<z<$17), Epoch of X-ray Heating (EoX; 17$<z<$10), and the Epoch of Reionisation (EoR; 10$<z<$5.3), none have the sensitivity or resolution to map the structure of the hydrogen through this period, or to deeply explore the Cosmic Dawn (\citenum{mertens20,barry19a,jacobs16,trott20,hera21}). SKA1-Low's collecting area and angular resolution will allow for the exquisite calibration and subtraction, and deep integration of signal, required to address this challenge (\citenum{barry16,patil16,trottwayth2016}).

Along with this capability, comes the reality of a complex, multi-component instrument and signal chain, which present significant data treatment and analysis challenges. The large collecting area at low radio frequencies necessitates the use of aperture arrays of receptor elements to form stations. SKA1-Low stations will be comprised of 256 wide-band dual-polarisation dipole antennas, spread over a diameter of 38~metres, with a minimum spacing of $\sim$1.5~m. Sub-arrays of 16 dipoles will be aggregated at the station before those signals are beamformed (\citenum{eloy20}). The Aperture Array Verification System 2.0 (AAVS, \citenum{sokolowski21,Macario2021}) is a test station of SKALA4 dipoles located at the Murchison Radioastronomy Observatory (MRO). It is designed to be a full SKA1-Low prototype station for system testing and calibration. Its dipole configuration will be used in this work.

Each station's dipole configuration can be varied. Initially, SKA1-Low's 512 stations were intended to have unique configurations. This design allows for residual sidelobe structure to be reduced by incoherent combination of different and random sidelobes from the array factor component of the instrument (the individual dipole locations that imprint structure on the station voltage beam pattern). This was studied in detail by Mort et al (\citenum{mort17}), who also explored apodization as a method for controlling sidelobes, at the expense of sensitivity. In reality, the station beam model for a single configuration is complicated, with sophisticated Fully Embedded-Element (FEE) simulations required at all frequencies, to compute the actual response to the sky, due to the complexity of the dipoles and the coupling of their signals due to their close spacing. This computational load, and the need to have an exquisite understanding of the behaviour of the complete instrument, motivates consideration of a much smaller number of unique station configurations (1--10) that can be accurately modelled, but can also provide some sidelobe suppression. In this work, the FEE simulations are not available (initial work on this is described in \citenum{Bolli2021,bolli20}), and we employ a simpler model where individual frequency-dependent element patterns are used within the station array factor, thereby neglecting cross-coupling effects between dipoles (\citenum{sokolowski21}). The unique configurations are generated via simple rotations of the AAVS dipole configuration, similar to the approach taken by the LOFAR telescope (\citenum{vanhaarlem13}) high-band antenna stations. Unlike LOFAR, where 48 antennas are spread over the station, SKA has more closely-packed antennas, where station rotation does not translate to a direct beam rotation, due to the complex mutual coupling between elements. We expect that the results here can be taken as the best-case scenarios.

Accurate and complete sky models are crucial for calibration of non-redundant low-frequency arrays, where large fields-of-view, complicated beam patterns, and cosmological signal weakness require that the flux is correctly accounted for to form the calibration sky model (\citenum{barry16,trottwayth2016,patil16}). Sardarabadi \& Koopmans (\citenum{sardarabadi19}) explored the calibratability of LOFAR and SKA1-Low under different model assumptions using a Fisher analysis approach, focussing on array layout rather than station configuration considerations. Nonetheless, they demonstrated calibratability of the full array under certain assumptions.

For current EoR experiments, there are two major elements for success that are relevant to this work: (1) overall sidelobe level, in order to control the power from sources outside of the field-of-view; (2) residual instrumental spectral structure (chromaticity), which dictates the leakage of source power into regions of parameter space that are used for science. Redundant arrays can be calibrated without as much reliance on a sky model, but it is still required to set some calibration parameters (e.g., \citenum{byrne19,joseph18}). Departures from redundancy, which occur in all real systems, degrade the ideal, and sky model information is required. Such a hybrid approach aims to remain sufficiently spectrally smooth to not affect the EoR signal. Current EoR experiments either subtract as many sources from the data as possible (e.g., \citenum{jelic08,chapman12,jacobs16}) to reduce contamination, or rely on excellent calibration and an achromatic instrument response to contain the foreground power in one part of the power spectrum parameter space (\citenum{hera21}). Even for the LOFAR telescope (\citenum{vanhaarlem13}), which currently subtracts the most sources from its data, sources beyond the second sidelobe are rarely removed due to their weakness and the lack of sufficient understanding of the station beam (\citenum{yatawatta13}). The suppression by the LOFAR beam away from phase centre means that sources do not need to be subtracted to the same depth.

This work is placed in the context of existing studies of SKA station calibration. Wijnholds and Bregman [\citenum{wijnholds2014}] studied the calibratability of SKA1-Low stations, including discussion of classical and sidelobe confusion in the context of station and array configurations. Sokolowski et al (\citenum{sokolowski21}) measured the sensitivity performance of AAVS, while van Es et al (\citenum{vanes}) discussed its calibration, and Mort et al (\citenum{mort17}) studied imaging performance for randomised station configurations. In this work, we expand on those with specific simulations to study the effect on EoR/CD experiments, with a view to exploring different numbers of unique configurations. We do not consider station apodization, but retain full sensitivity of the station to study the intrinsic source spectral behaviour.

\section{Methods}
\subsection{Power spectrum}
The SKA EoR/CD experiments aim to observe spatial fluctuations in the 21~cm brightness temperature over $z$=5.5--27 (55--240~MHz, \citenum{koopmans15}). They will comprise multi-redshift spatial power spectra ($z$=5.5--27), tomographic imaging ($z$=5.5--12), and 21~cm Forest probes along lines-of-sight to high-redshift radio loud quasars (\citenum{mellema13,koopmans15}). Tomographic imaging and 21cm Forest will be new and unique to SKA, with EoR and CD power spectra being pursued by current experiments. In this work, we focus on power spectra across the full frequency range of interest.

In its simplest form, the power spectrum is computed from the Fourier representation of the temperature fluctuations, $T(k)$, within an observational volume, $V=\Omega \Delta\nu$ Mpc$^3$, where $\Omega$ is the field-of-view and $\Delta\nu$ is the bandwidth (converted to cosmological units), and $k=|\bf{k}|$ denotes the spatial wavenumber (inverse Mpc)\footnote{We define our Fourier convention as $\tilde{T}({\bf k}) = \int_\Omega T({\bf r}) \exp{2\pi{j} {\bf k}\cdot{\bf r}}d{\bf r}$}. The dimensionless spherically-averaged (1D) power spectrum is defined by:
\begin{equation}
    \Delta^2(k) = \frac{k^3 P_k}{2\pi^2} \,\, {\rm mK}^2,
\end{equation}
with the power spectrum formed via,
\begin{equation}
    P(k) = \frac{\langle T(k) T^\ast(k) \rangle}{V} \,\, {\rm mK^2 Mpc^3},
\end{equation}
where the ensemble average runs over different realisations for that wavenumber. An interferometer measures the coherence of the specific intensity between two receptors at a given frequency, with the visibility measurement at mode vector $(u,v,w)$ and frequency $\nu$ defined as:
\begin{equation}
    \tilde{S}(u,v,w;\nu) = \int_\Omega I(l,m;\nu)B(l,m;\nu) \exp{-2\pi{j}(ul+vm+w(\sqrt{1-l^2-m^2}-1))} \frac{dldm}{\sqrt{1-l^2-m^2}},\,{\text Jy}
\end{equation}
where $B(l,m;\nu)$ is the primary beam attenuation factor formed from the voltage beams $D$ of the two receptors forming the baseline,
\begin{equation}
    B(l,m;\nu) = D_1(l,m;\nu) D_2^\ast(l,m;\nu).
\end{equation}
For many arrays, the voltage beams are identical, but in general they can vary. For SKA1-Low, there will be 1--10 different station configurations. In the flat-sky approximation, where the field-of-view is small and the baselines short, the measurement equation reduces to a 2D Fourier transform,
\begin{equation}
    \tilde{S}(u,v;\nu) \simeq \int_\Omega I(l,m;\nu)B(l,m;\nu) \exp{-2\pi{j}(ul+vm)} dldm.
\end{equation}
The specific intensity, $I$ is related to the brightness temperature through,
\begin{equation}
    I = \frac{2k_BT}{\lambda^2},
\end{equation}
where $k_B$ is Boltzmann's constant. The measured visibilities therefore encode the angular Fourier transform of a quantity that is proportional to the sky temperature, thereby performing part of the task required to compute the power spectrum.

The final line-of-sight Fourier transform along the frequency axis encodes the line-of-sight wavenumber, $k_\parallel$. As is standard, this is performed with a window function, $\Upsilon(\nu)$ (here, a 7-term Blackman-Harris, (\citenum{SASPWEB2011})), to reduce the spectral leakage caused by a finite bandwidth, such that the transform from $\nu$ to $\eta$ is:
\begin{equation}
    \tilde{S}(\eta;u,v) = \int_{BW} \Upsilon(\nu) \tilde{S}(\nu;u,v) \exp{(-2\pi{j}\nu\eta)} d\nu.
\end{equation}
The 7-term Blackman-Harris window yields high sidelobe suppression, at the expense of a 4-cell main lobe. After transform, the data are in the full wavenumber space, with $(u,v,\eta)$ linearly proportional to cosmological wavenumbers, $(k_x,k_y,k_z)$ according to the transforms in Ref. \citenum{morales04}, in the ergodic limit.

Despite the spherically-averaged power spectrum encoding complete spatial variance information about the isotropic 21~cm signal, it is often used only after study of the cylindrically-averaged (2D) power spectrum. The latter estimates power as a function of angular and line-of-sight scales separately ($k_\bot,k_\parallel$). In this space, the isotropic 21~cm signal imprints symmetric power, but contaminating signal from intervening continuum radio galaxies and our own Galaxy imprint anisotropic power, due to their flat spectral response. The line-of-sight Fourier transform concentrates continuum power at low $k$ modes (the `foreground wedge'), keeping higher modes clean for exploration of the cosmological signal (the `EoR window'). However, the realities of fixed bandwidth, calibration errors and instrumental chromaticity due to primary beam spectral structure result in power being leaked from the wedge to the window, yielding systematics-limited results. In this work we focus on the sources located far from the pointing centre, where the beam is poorly-understood, has spatial and spectral structure, and sources are typically too faint to self-calibrate and peel from the dataset. We study the imprint of these sources on the EoR and CD power spectra for various SKA station configurations.

\subsection{SKA1-Low configurations}
We use the current plans for station array layout for SKA1-Low to provide the most faithful representation of the instrument model (\citenum{dewdney16}), and the dipole layout for the Aperture Array Verification System 2.0 (AAVS2), deployed at the Murchison Radioastronomy Observatory (MRO). This base layout provides one configuration of dipoles, and others are formed via a 45$N$ degree rotation of the dipole locations in the station, where $N=1,2,...,8$. Example configurations are displayed in the Appendix. Dipole element patterns were obtained from the LFAA consortium, with measurements of the full-sky pattern for both X and Y polarisations at 1~MHz intervals over a 50--300~MHz band, spline interpolated to 200~kHz.

The zenith station beam is formed via placement of these element patterns, $E$, at the actual AAVS2 dipole positions, and combined with the array factor at wavelength $\lambda$:
\begin{equation}
    D^X(l,m;\lambda) = \displaystyle\sum_{i=1}^{256} E_i(l,m) \exp{-2\pi{j}(u_il+v_im)},
\end{equation}
where $u_i = (x_i-x_{\rm ref})/\lambda$ is the dipole position wavenumber relative to an arbitrary dipole. Off-zenith beams are formed via a translation of the phase centre, $(l-l_{\rm ref},m-m_{\rm ref})$. The Stokes I beam is then formed via the average of the two auto-polarisations from each station:
\begin{equation}
    B(l,m) = \frac{D_1^{\ast X} D_2^X + D_1^{\ast Y} D_2^Y}{2}.
\end{equation}

\begin{figure}
\begin{center}
\begin{tabular}{c}
\includegraphics[height=14.5cm]{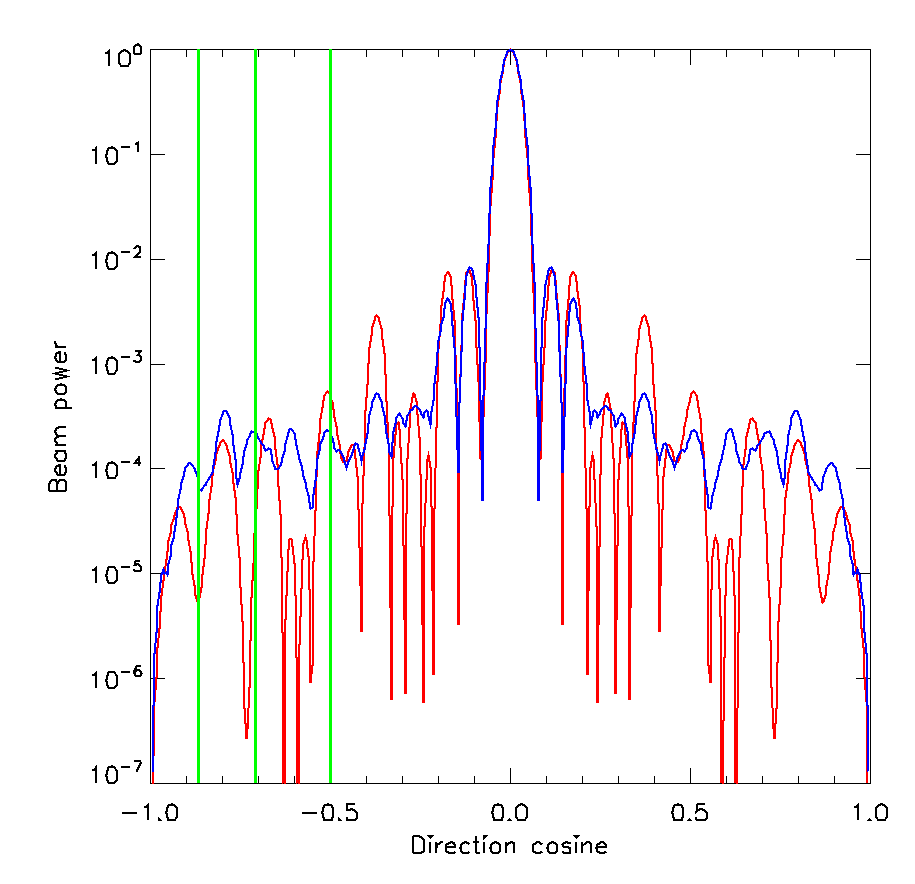}
\end{tabular}
\end{center}
\caption 
{ \label{fig:beams}
Cut through example beam powers for $N_b=4$ (blue), and $N_b=1$ (red). $N_b=4$ shows lower sidelobe level than the single configuration at some angles. Vertical green lines delineate regions considered in this work, $\theta_Z >$ 30, 45, 60 degrees from zenith. } 
\end{figure}

Figure \ref{fig:beams} shows a cut through two baseline beams (power, formed via cross-multiplication of station voltage patterns) for a single station configuration ($N_b=1$, red) and four unique configurations ($N_b=4$, blue) at 60~MHz. The latter shows clear sidelobe suppression at some angles, but retains complex structure. Vertical green lines delineate regions considered in this work, $\theta_Z >$ 30, 45, 60 degrees from zenith.

With identical station dipole configurations, spatial and spectral structure within the polarised voltage beams (due to their composition from 256 individual elements) combine coherently in the baseline beam formed from cross-correlating two stations. Measurements formed from baselines with stations having differing layouts will have a different structure, which may have lower far sidelobes if the voltage beams destructively interfere.

\subsection{Sky model}
We form the sky model from a point source population only, thereby representing the minimum contamination case where complex, extended structures are omitted. The model differential source counts is constructed from a parametric fit to observations at 150~MHz (\citenum{franzen,intema,lynch21}):
\begin{equation}
    \frac{dN}{dS} = \alpha (S/S_0)^\beta \,\, {\rm Jy^{-1}sr^{-1}}, \label{eqn:dnds}
\end{equation}
where $\alpha=4100\, {\rm Jy^{-1}sr^{-1}}$, $\beta=-1.59$ for $S<1$~Jy and $S_0=1$~Jy and, steepening to $\beta=-2.5$ for $S>1$~Jy. Sources are distributed randomly in the sky beyond a pre-defined angle from zenith, in order to study the effect of sources outside of the main lobe. For an inner angle of 45 degrees from zenith, there are $\sim$750,000 sources with intrinsic flux densities between 1~mJy and 5~Jy.

\subsection{Simulations}
We simulate noise-free visibilities from flat-spectrum sources that are generated and placed randomly in the sky beyond a zenith angle of $\theta_Z$ and attenuated by a Stokes I beam. Each array configuration uses 1--8 unique voltage beams, which are constructed from rotations of the AAVS dipole locations (not rotations of the beams; polarisation position angle is preserved) at 45 degree intervals. A synthesis of 1.0 hour spanning zenith is simulated to gain the benefits of sources transiting through beam sidelobes (where 1~hour is sufficient for a source to pass through a full sidelobe).

Each station in the array is assigned a random dipole configuration from the 1--8 configurations, and those used for forming the beam for each baseline. Source positions and flux densities are generated according to Equation \ref{eqn:dnds} for $\theta > \theta_Z$, attenuated by a beam model, and gridded onto a model sky for each 200~kHz channel. The beam for a given baseline is formed from the voltage patterns for the two stations, and the $uv$-plane is generated for each configuration combination via an FFT. Measured visibilities are then extracted from the $uv$-planes at the locations appropriate for the pointed array, using a spline interpolation from the grid in the real and imaginary components. Typically, in an EoR experiment these `measured' visibilities would then be gridded onto a measurement $uv$-plane with a beam kernel, to build the $u,v,\nu$ cube of data for further analysis. For the purposes of this work, where we want to explore chromaticity from the beam, we instead directly deal with the measured visibilities via a delay transform approach (\citenum{parsons10,parsons12}). This approach removes the possibility of gridding artifacts introducing spectral structure, and allows us to focus on the beam influence. In this case, the final transform from frequency to $\eta$ is performed along each visibility individually.

\begin{figure}
\begin{center}
\begin{tabular}{c}
\includegraphics[height=14.5cm]{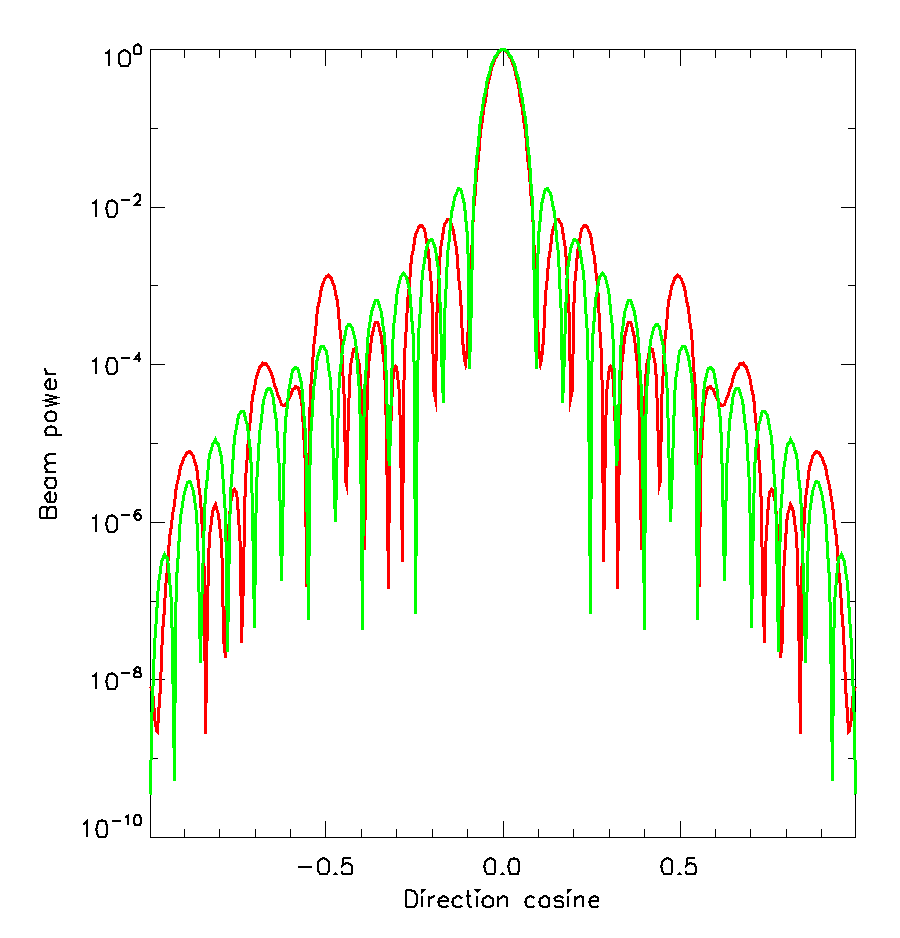}
\end{tabular}
\end{center}
\caption 
{ Cut through beam power (voltage squared) for the AAVS2 configuration (red) and an achromatic matched-width Airy disk (green) for $z=19.5$. \label{fig:airy}} 
\end{figure}

In addition to the AAVS2 station layout with element pattern, we simulate a simple Airy disk with a width to match the main lobe width from AAVS2, in an effort to simulate observations with a filled dish aperture. The Airy disk profile, generated via a jinc function, is attenuated by the same element pattern to simulate a smooth aperture. The beam voltage pattern, for both X and Y polarisations is given by:
\begin{equation}
    D^A(l,m) = \frac{J_1(r)}{r}E(l,m),
\end{equation}
where $r = 2\pi{a}\sqrt{l^2+m^2}/\lambda$, and $a$ is the aperture radius for a Bessel function of the first kind, $J_1(x)$. The Airy disk beam is kept achromatic to test the degree to which beam chromaticity imprints power. The Airy disk is applied to one experimental configuration only for comparison (60--80~MHz, $\theta_z>$~45 degrees). A cut through the Airy disk beam power (green) and the AAVS2 configuration (red) is shown in Figure \ref{fig:airy} for $\nu$=60~MHz. Overall the sidelobe level is similar between the two, with more structure observed for the latter, however there are sidelobes where the difference between the two is almost one dex, which has the potential to contribute a two dex power differential in the power spectra (beam power squared).

Table \ref{table:sims} shows the different simulations performed, where the angle cut, $\theta_Z$, and the number of unique station configurations, $N_b$, are varied, across three redshifts. In all cases, the observation bandwidth is fixed to 20~MHz, corresponding to $\sim$10~MHz effective bandwidth with the Blackman-Harris window function, the minimum and maximum intrinsic source flux densities are 1~mJy and S$_{\rm max}$=5~Jy. We also explore an upper limit of S$_{\rm max}$=1~Jy in some cases to demonstrate the impact of deeper cleaning. The brighter sources dominate the signal budget and are the most critical to leaked power.
\begin{table}
\centering
\begin{tabular}{|c||c|c|c|c|c|}
\hline 
$\nu$ (MHz) & $z$ & $N_b$ & $\theta_Z$ (deg.) & $N_{\rm src}$ & S$_{\rm max}$ \\ 
\hline \hline 
60--80 & 19.5 & 1 & 45 & 750k & 5~Jy\\
60--80 & 19.5 & 2 & 45 & 750k & 5~Jy\\
60--80$^{\ast}$ & 19.5 & 4 & 45 & 750k & 5~Jy\\
60--80 & 19.5 & 8 & 45 & 750k & 5~Jy\\
60--80 & 19.5 & 4 & 60 & 355k & 5~Jy\\
60--80 & 19.5 & 8 & 60 & 338k & 1~Jy\\
60--80 & 19.5 & 4 & 30 & 1280k & 5~Jy\\
\hline
100--120 & 11.5 & 1 & 45 & 750k & 5~Jy\\
100--120 & 11.5 & 2 & 45 & 750k & 5~Jy\\
100--120 & 11.5 & 4 & 45 & 750k & 5~Jy\\
100--120 & 11.5 & 8 & 45 & 750k & 5~Jy\\
100--120 & 11.5 & 4 & 60 & 355k & 5~Jy\\
100--120 & 11.5 & 8 & 60 & 338k & 1~Jy\\
100--120 & 11.5 & 4 & 30 & 1280k & 5~Jy\\
\hline
140--160 & 8.5 & 1 & 45 & 750k & 5~Jy\\
140--160 & 8.5 & 2 & 45 & 750k & 5~Jy\\
140--160 & 8.5 & 4 & 45 & 750k & 5~Jy\\
140--160 & 8.5 & 8 & 45 & 750k & 5~Jy\\
140--160 & 8.5 & 4 & 60 & 355k & 5~Jy\\
140--160 & 8.5 & 8 & 60 & 338k & 1~Jy\\
140--160 & 8.5 & 4 & 30 & 1280k & 5~Jy\\
\hline
\hline
\end{tabular}
\vspace{0.1cm}
\caption{Different sky and telescope configurations used for different simulations of the power spectrum. $^{\ast}$ This configuration is also used for the Airy disk comparison beam.}\label{table:sims}
\end{table} 
In the table we explore the dimensions of angle cut and number of unique stations, separately, assuming $N_b=4$ for testing different angle cuts.

To compare the point source power to a model for the cosmological signal, we use 21cmFAST (\citenum{mesinger11}) output power spectra at the relevant redshift, and assume isotropy to deproject the signal power to 2D.

\section{Results}
\subsection{Cosmic Dawn; $z$=19.5, 60--80~MHz}
\begin{figure}
\begin{center}
\begin{tabular}{c}
\includegraphics[height=18.cm]{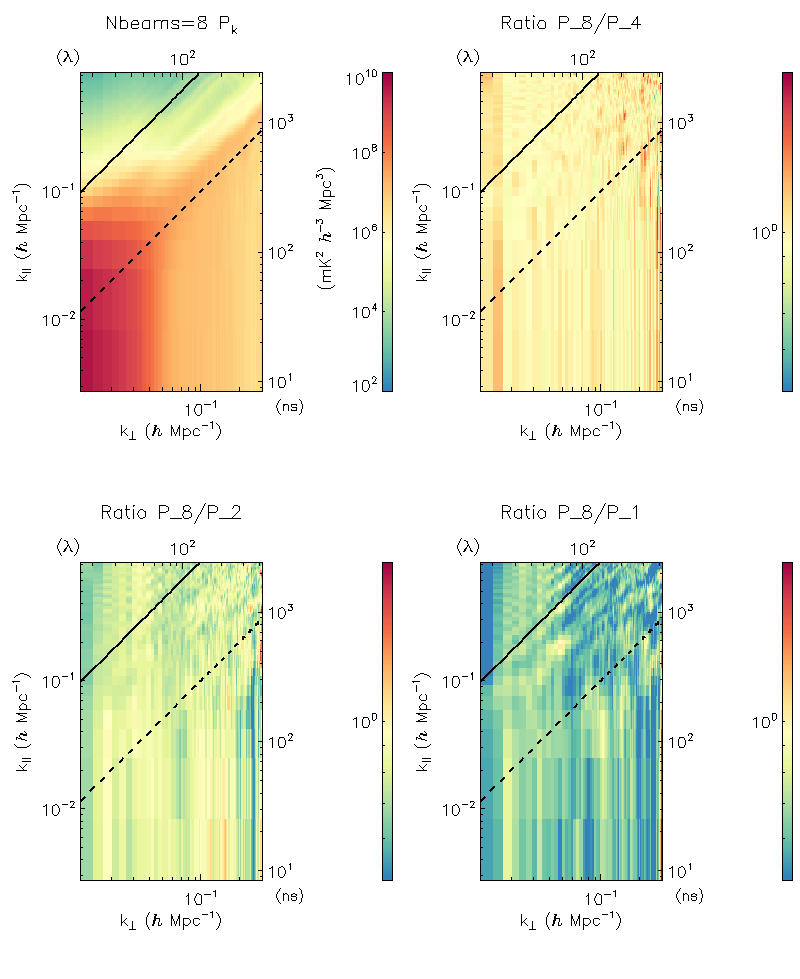}
\end{tabular}
\end{center}
\caption 
{ \label{fig:CD_4plot}
$z=19.5$: 2D power spectra of sources beyond $\theta_Z=45$ degrees for $N_b=8$ (top-left), and ratios of $N_b=8$ power spectra to $N_b=4$ (top-right), $N_b=2$ (bottom-left), $N_b=1$ (bottom-right), at $z=19.5$. $N_b=4,8$ both show significantly less power than those with a single or two unique configurations. The dashed black line denotes an approximate first beam null line, while the solid black line denotes an approximate horizon scale. } 
\end{figure}

At high redshifts, we expect the signal to be accessible only via power spectra, with tomography posing a challenge. We consider
the performance of the telescope, with a zenith-phased observation and one-hour rotation synthesis. Figure \ref{fig:CD_4plot} shows the 2D power spectra for $N_b=8$, $\theta_Z>$ 45 degrees (top-left), and ratios of $N_b=8$ power spectra to $N_b=4$ (top-right), $N_b=2$ (bottom-left), $N_b=1$ (bottom-right). $N_b=4,8$ both show significantly less power than those with a single or two unique configurations. The power spectra show the expected characteristics of an interferometer; a foreground block at low $k_\parallel$ that extends diagonally into higher $k_\parallel$ modes due to baseline migration (mode-mixing; the `wedge'), beam sidelobes parallel to the diagonal edge of the wedge, and a cleaner `EoR window' beyond the sidelobes at higher $k_\parallel$.

\begin{figure}
\begin{center}
\begin{tabular}{c}
\includegraphics[height=6.0cm]{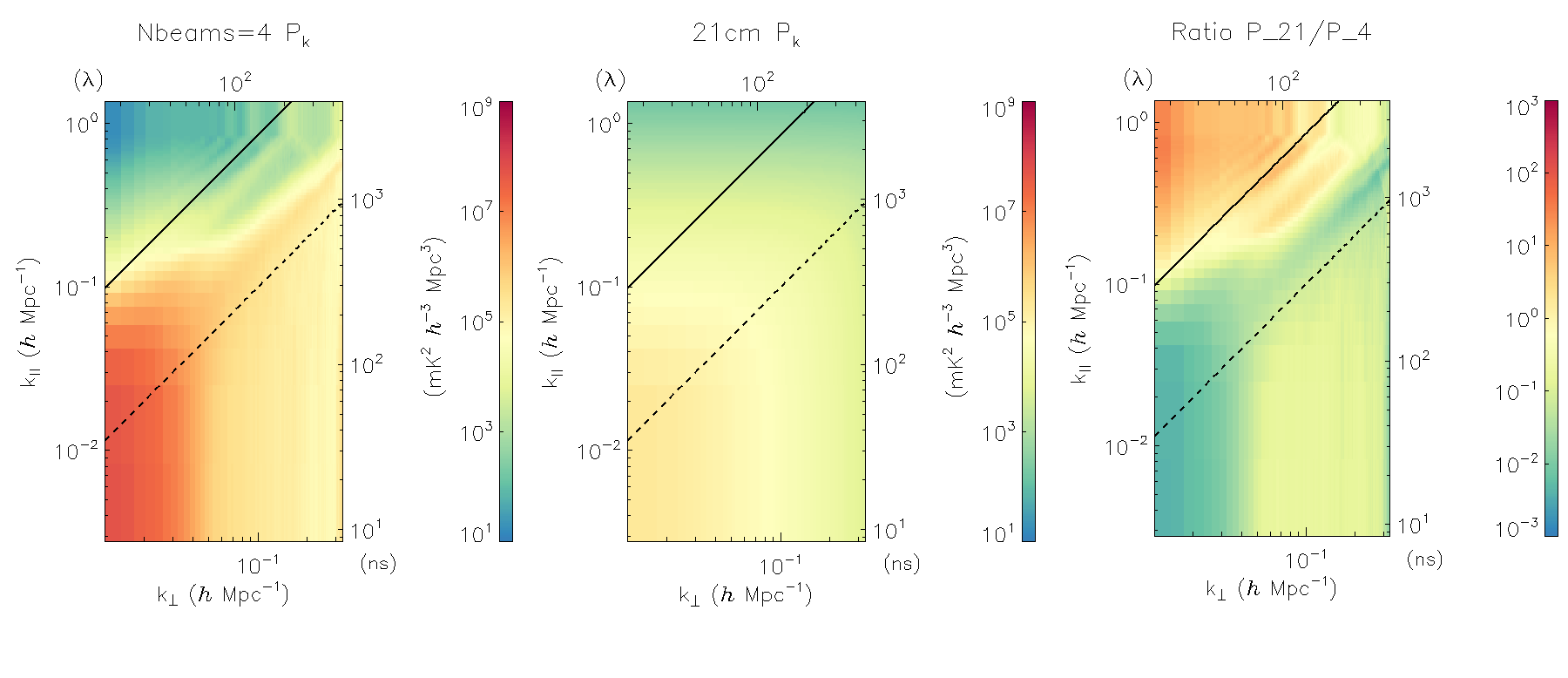}
\end{tabular}
\end{center}
\caption 
{ Power ($N_b=4$, left), typical 21cm signal (middle) and ratio (right) for $z=19.5$ and $\theta_Z>60$ degrees. \label{fig:CD_3plot}} 
\end{figure}

\begin{figure}
\begin{center}
\begin{tabular}{c}
\includegraphics[height=6.5cm]{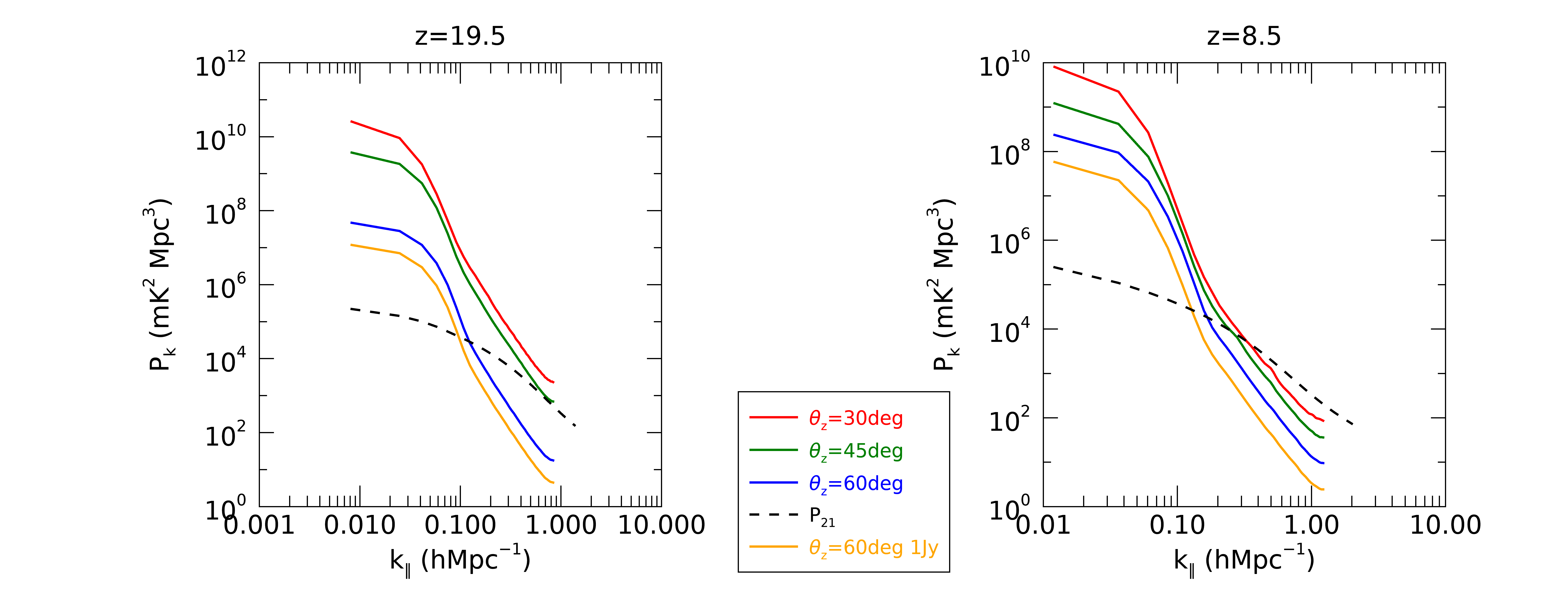}
\end{tabular}
\end{center}
\caption 
{ \label{fig:angles}
(Left) Power versus $k_\parallel$ at $z=19.5$ for $N_b=4$ and $\theta_Z>$ 30 (red), 45 (green) and 60 (blue [orange]) degrees for $k_\bot=0.01 h$Mpc$^{-1}$ and S$_{\rm max}$~=~5~Jy [1~Jy], and a comparison 21~cm power spectrum (black-dashed). (Right) Same but for $z=8.5$. } 
\end{figure}

The spectra show that most of the power is removed from the point source wedge, consistent with larger suppression of sources for more configurations. However, there is also a deficit in the EoR window for $N_b=4,8$ compared with $N_b=2,1$, suggesting that the leaked signal power is also reduced when there are more unique configurations. Figure \ref{fig:CD_3plot} shows a $N_b=4$ $\theta_Z>60$ degrees noise-free power spectrum (left), with a comparison 21cmFAST signal spectrum on the same colour-scale (middle), and the ratio of the two (right). With no noise in the simulations, the ratio should tend to infinity in the EoR Window if there was no contamination.
Signal in the foreground wedge is inaccessible due to the point source power, while the SNR in the EoR window exceeds unity for several angular modes. The same analysis can be undertaken for $\theta_Z>$ 30, 45 degrees, and for both maximum flux densities. Figure \ref{fig:angles} (left) shows a cut through the 2D spectrum at $k_\bot=0.01 h$Mpc$^{-1}$ as a function of $k_\parallel$ and the four point source angle cuts and S$_{\rm max}$ combinations (solid lines), and the comparison 21cmFAST spectrum (black-dashed). More modes are accessible when sources are removed closer to the horizon, with $\theta_Z>$ 60 degrees and S$_{\rm max}$=1~Jy preferred for accessing a large amount of parameter space.

\begin{figure}
\begin{center}
\begin{tabular}{c}
\includegraphics[height=6.5cm]{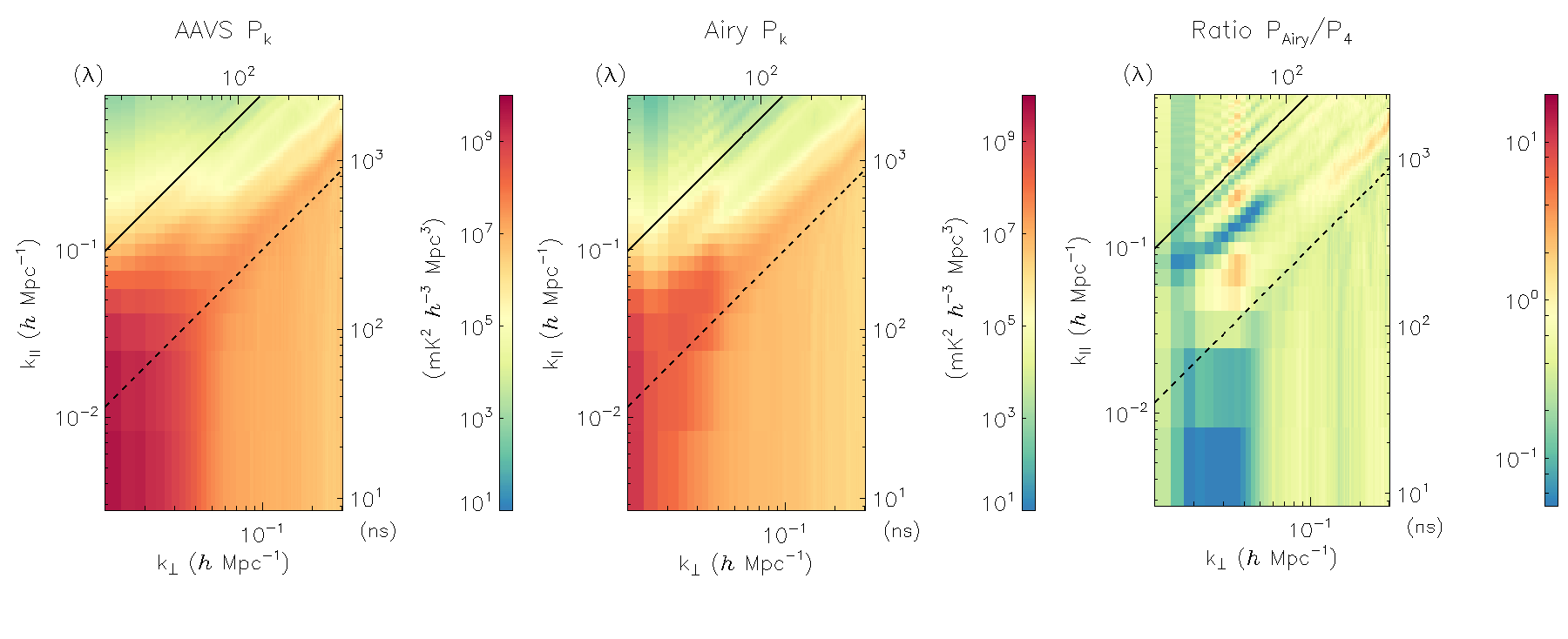}
\end{tabular}
\end{center}
\caption 
{Power spectra for the same experimental setup at $z=19.5$ for $N_b=4$ (left), the matched Airy disk (centre), and the ratio of the two (right). \label{fig:airyaavs}
} 
\end{figure}

For comparison, we apply the $\theta_Z>$ 45 degrees simulation to the achromatic Airy disk displayed in Figure \ref{fig:airy}. For clarity, only the Airy disk is achromatic, and all other factors in the experiment are identical to that for AAVS ($uv$-coverage, chromatic instrument response, rotation synthesis etc.). Figure \ref{fig:airyaavs} shows power spectra for the same experimental setup at $z=19.5$ for $N_b=4$ (left), the matched Airy disk (centre), and the ratio of the two (right).
There are two clear results from the ratio plot. Firstly, the foreground Airy disk power shows a moderate overall reduction in level across much of the parameter space, and therefore less power in the measurements. Secondly, the EoR window shows similar power levels to AAVS, demonstrating that spectral leakage due to beam chromaticity is not playing a measurable role for the AAVS aperture array.
 Despite the Airy disk having significant angular sidelobe structure, it is designed to have no spectral structure to demonstrate the importance of this feature.

\subsection{Epoch of X-ray Heating; $z$=11.5, 100--120~MHz}
\begin{figure}
\begin{center}
\begin{tabular}{c}
\includegraphics[height=18.cm]{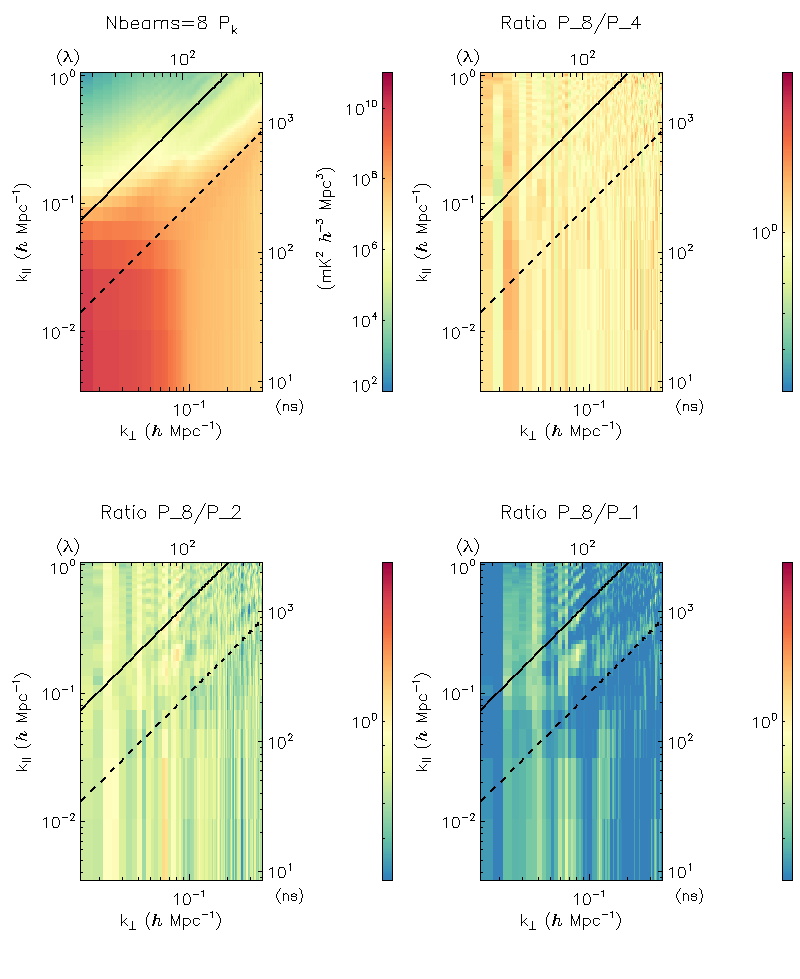}
\end{tabular}
\end{center}
\caption 
{ \label{fig:EoX_4plot}
$z=11.5$: 2D power spectra for sources beyond $\theta_Z=45$ degrees for $N_b=8$ (top-left), and ratios of $N_b=8$ power spectra to $N_b=4$ (top-right), $N_b=2$ (bottom-left), $N_b=1$ (bottom-right), at $z=11.5$. $N_b=4,8$ both show significantly less power than those with a single or two unique configurations.} 
\end{figure}

\begin{figure}
\begin{center}
\begin{tabular}{c}
\includegraphics[height=5.5cm]{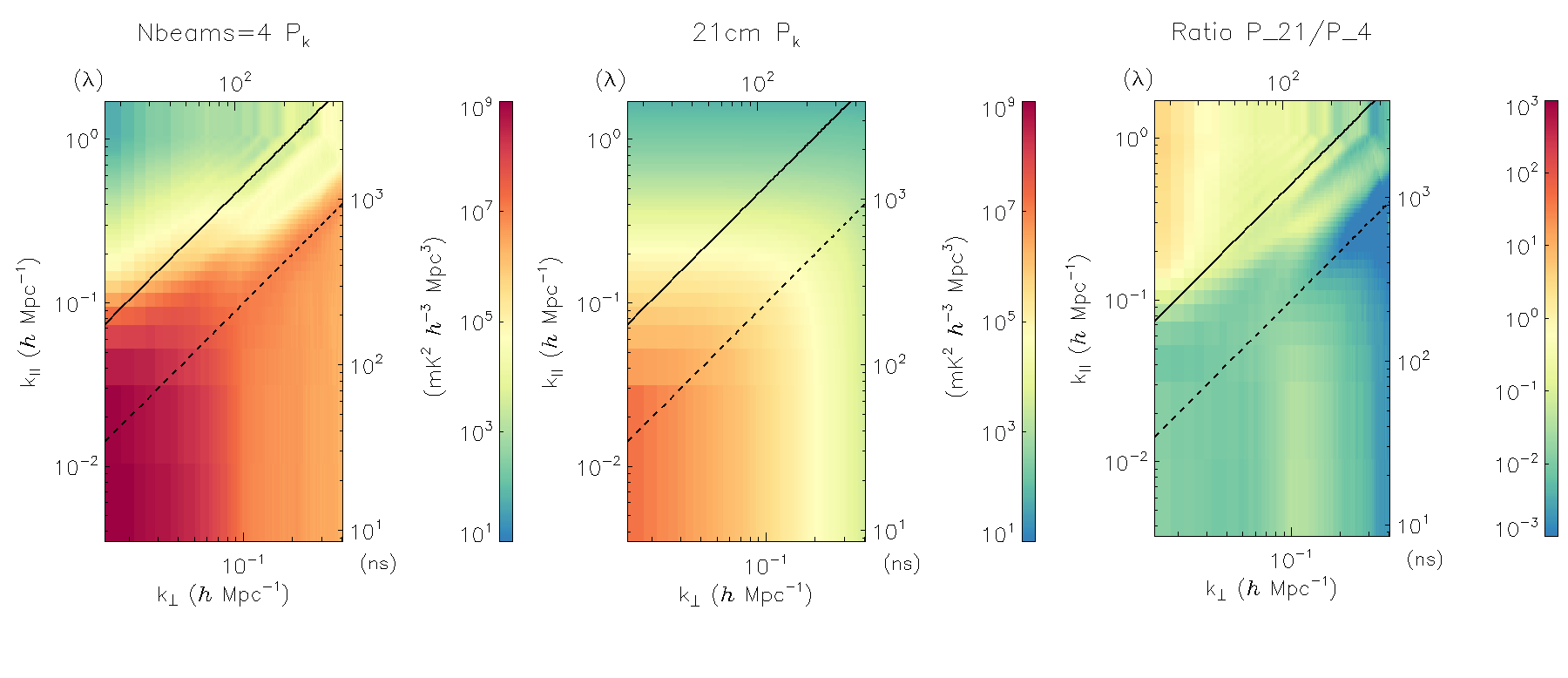}
\end{tabular}
\end{center}
\caption 
{ Power ($N_b=4$, left), typical 21cm signal (middle) and ratio (right) for $z=11.5$ and $\theta_Z>60$ degrees, S$_{\rm max}$=5~Jy. \label{fig:EoX_plot3}} 
\end{figure}

We expect the FM radio band from $\sim$80--95~MHz to be corrupted by reflected signals from Geraldton and Perth, as found by Wilensky et al. (\citenum{wilensky19}), making observations in the Epoch of X-ray Heating challenging. We instead study the end of the EoX at $z=11.5$. Figure \ref{fig:EoX_4plot} shows the 2D power spectra for $N_b=8$, and ratios to that, for $\theta_Z>$ 45 degrees, and including rotation synthesis.
Here, like in the Cosmic Dawn, $N_b=8,4$ perform comparably to each other, but with improved performance over $N_b=1,2$. The $N_b=4$ and $N_b=8$ configuration ratio is close to unity, but with angular structure corresponding to their different sidelobe distribution being observed at different $k_\bot \propto |u|$ modes. The ratio to the 21cmFAST signal power in Figure \ref{fig:EoX_plot3} (right) shows that accessing modes within the foreground wedge will require point sources to be removed beyond $\theta_Z=$ 60 degrees.

\subsection{Epoch of Reionisation; $z=8.5$, 140--160~MHz}
The EoR will be explored via power spectrum and tomography, and is expected to be the most straightforward of the epochs to observe. In order to exceed any detections from the current generation of experiments probing this era (MWA, LOFAR, HERA), SKA would be aiming to produce deep observations over a large range of angular and line-of-sight modes.

\begin{figure}
\begin{center}
\begin{tabular}{c}
\includegraphics[height=18.cm]{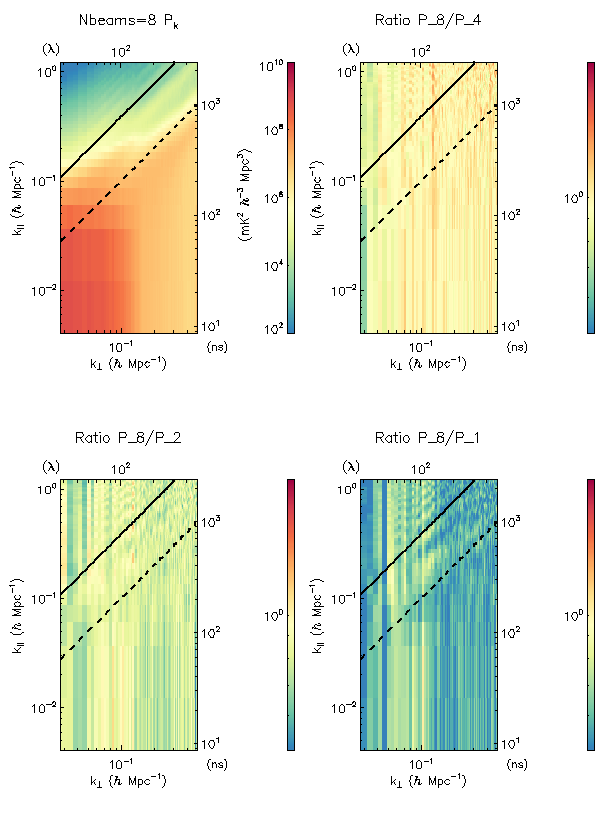}
\end{tabular}
\end{center}
\caption 
{ \label{fig:EoR_4plot}
$z=8.5$: 2D power spectra of sources beyond $\theta_Z=45$ degrees for $N_b=8$ (top-left), and ratios of $N_b=8$ power spectra to $N_b=4$ (top-right), $N_b=2$ (bottom-left), $N_b=1$ (bottom-right). $N_b=4,8$ both show significantly less power than those with a single or two unique configurations.} 
\end{figure}

\begin{figure}
\begin{center}
\begin{tabular}{c}
\includegraphics[height=6.5cm]{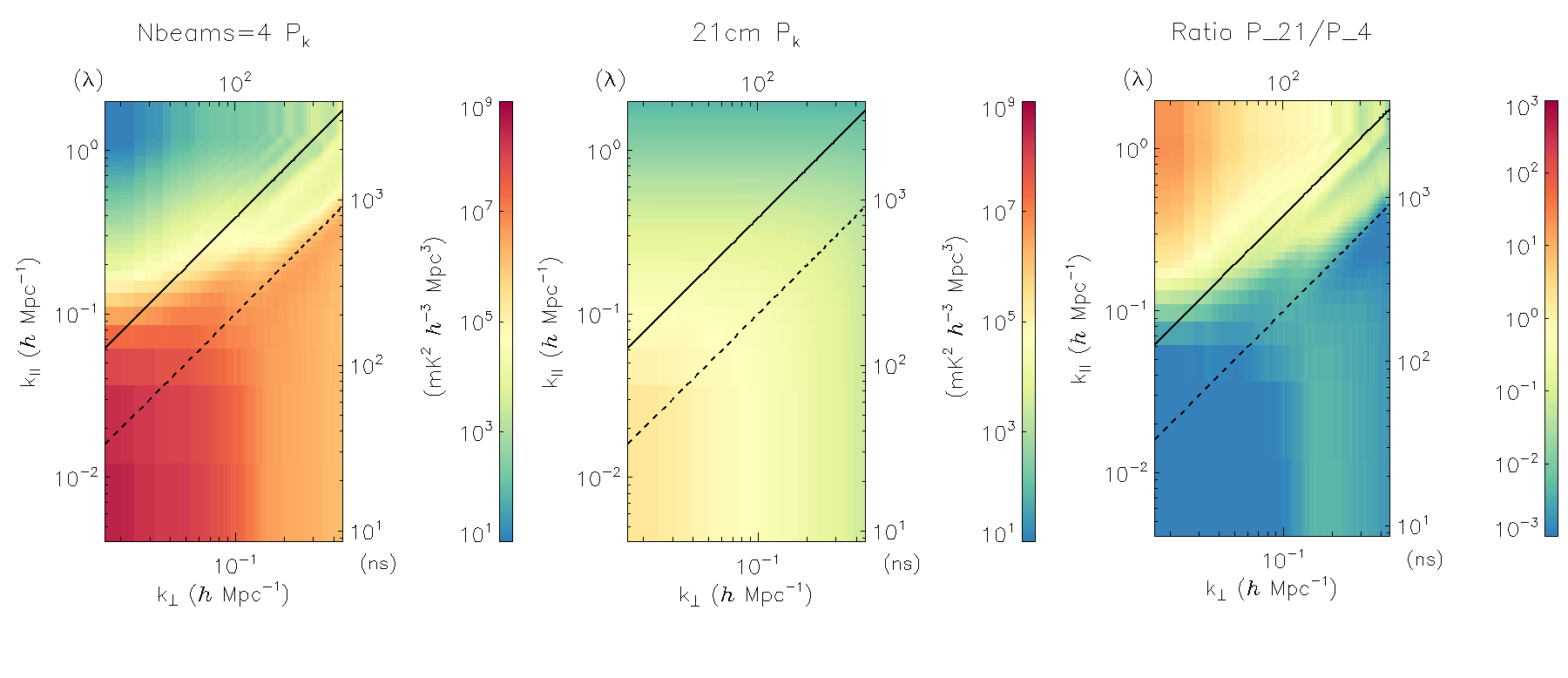}
\end{tabular}
\end{center}
\caption 
{ Power ($N_b=4$, left), typical 21cm signal (middle) and ratio (right) for $z=8.5$ and $\theta_Z>60$ degrees. \label{fig:EoR_plot3}} 
\end{figure}

Figure \ref{fig:EoR_4plot} shows the same spectra as for previous redshifts; $N_b=8$ and ratios to that, for $\theta_z>$~45 degrees.
As observed for $z=11.5$, the performance with more than two unique configurations is improved over a single or two configurations. $N_b=4,8$ concentrate power in the foreground wedge, with better performance in the EoR Window. Figure \ref{fig:angles} (right) shows a cut through $k_\bot=0.01 h$Mpc$^{-1}$ for $N_b=4$ and the three levels of source zenith angle cut. Like for $z=19.5$ there is clear improvement for different levels of source cut and source subtraction depth, showing that this is a contributor to the excess power. For this redshift more than others, the relative weakness of the cosmological signal and the amplitude of the beam sidelobes, mean that residual source power is more dominant in the power spectrum. The power ratio shown in Figure \ref{fig:EoR_plot3} (right) demonstrates that very few modes of the power spectrum parameter space have high SNR.

\section{Theoretical considerations}\label{sec:theory}
The results shown here via simulation can be motivated by theoretical considerations for point source power. In general, the root-mean-square integrated flux density of an ensemble of spatially-random sources is given by:
\begin{equation}
 S_{\rm rms} \approx \sqrt{N_{\rm src}}\langle S \rangle \,\, {\rm Jy}.
\end{equation}
Over a 10~MHz effective bandwidth, at $z=8.5$, and converting to cosmological units, this corresponds in the DC ($k=0$) mode to:

\begin{equation}
 P(k=0) \approx 9.5 \times 10^{11} \left(\sqrt{N_{\rm src}}\langle S \rangle \right)^2 \,\, {\rm mK^2 Mpc^3},
\end{equation}
for an unattenuated sky, and
\begin{equation}
 P(k=0) \approx 9.5 \times 10^{11} \left( \sqrt{N_{\rm src}}\langle S \rangle\langle B \rangle \right)^2\,\, {\rm mK^2 Mpc^3},
\end{equation}
observed. Figure \ref{fig:angles} demonstrates that the taper provides eight orders of magnitude of suppression from the DC mode to EoR modes, and those cosmological signals are in the range of $10^2-10^4$~mK$^2$~Mpc$^3$. As an example, for $z=8.5$ and $\theta_Z>$ 60 degrees, where $\sqrt{N_{\rm src}}\langle S \rangle \simeq 27.1$~Jy,
\begin{equation}
    P(k=1.0) \approx 7.0 \times 10^{6} \left(\langle B \rangle \right)^2\,\, {\rm mK^2 Mpc^3}
\end{equation}
on EoR modes, requiring 4--5 orders of magnitude beam suppression for revealing cosmological signal. The importance of the beam sidelobe level is readily apparent from this analysis. Beyond 60 degrees, the beam sidelobe level on a baseline is $\simeq$10$^{-3}$, corresponding to a suppression of $\simeq$10$^{-6}$ in power, yielding the $P_k \simeq 10^0$~mK$^2$~Mpc$^3$ at $k=1.0$ in Figure \ref{fig:angles}.

We can take the analysis further by understanding the dependence of $S_{\rm rms}$ on $S_{\rm max}$ and sky area, using Equation \ref{eqn:dnds}:
\begin{eqnarray}
    S_{\rm rms} &\approx& \sqrt{N_{\rm src}}\langle S \rangle = \frac{\displaystyle\sum S}{\sqrt{N_{\rm src}}}\,\,{\rm Jy},\\
    N_{\rm src} &=& \frac{\alpha\Omega}{1+\beta}\left(S_{\rm max}^{1+\beta} -  S_{\rm min}^{1+\beta} \right),\\
    \displaystyle\sum S &=& \frac{\alpha\Omega}{2+\beta}\left(S_{\rm max}^{2+\beta} -  S_{\rm min}^{2+\beta} \right),\\
    \rightarrow S_{\rm rms} &\simeq& \frac{\sqrt{\alpha\Omega(1+\beta)}}{2+\beta} \frac{S_{\rm max}^{2+\beta} -  S_{\rm min}^{2+\beta}}{\sqrt{S_{\rm max}^{1+\beta} -  S_{\rm min}^{1+\beta}}} \,\,{\rm Jy}.
\end{eqnarray}
With $S_{\rm max} \gg S_{\rm min}$, this reduces to,
\begin{equation}
    S_{\rm rms} \simeq \frac{\sqrt{\alpha\Omega(1+\beta)}}{2+\beta} \frac{S_{\rm max}^{2+\beta}}{\sqrt{S_{\rm max}^{1+\beta}}} = \frac{\sqrt{\alpha\Omega(1+\beta)}}{2+\beta} S_{\rm max}^{1.5+\beta/2} \,\,{\rm Jy}.
\end{equation}
The power spectrum therefore scales as:
\begin{equation}
    P(k) \propto \Omega S_{\rm max}^{3+\beta}.
\end{equation}
Hence, we expect that the power spectrum scales linearly with the zenith angle cut, and close to the square-root of the source subtraction limit. Peeling deeper therefore has a larger impact than peeling further from the zenith, with a reduction in power by a factor of two requiring cleaning of 1.6$\times$ deeper.

\section{Discussion and Conclusions}
This work presents an initial exploration of the impact of different station configuration choices for EoR/CD science with SKA1-Low. It also presents some options for how to undertake the experiment, and considerations that are important in its success. It is clear that at least 4 station configurations are important for reducing leakage. Experiments that require excellent instantaneous performance will not gain the benefits of synthesis, and chromaticity of sidelobes may play a larger factor. This is not the case for the EoR/CD experiments themselves, but will be important in the data calibration step, which is expected to be performed on a more frequent cadence (several minutes) than the time required for a source to transit a beam sidelobe (estimated to be $\sim$30 mins). This mismatch may result in calibration errors, which will propagate to the experimental data, but this effect is not considered here.

The results here consider two features: (1) overall sidelobe level, which dictates how much power remains in the foreground wedge; (2) residual beam chromaticity, which dictates the amount of power that leaks out of the wedge and into the EoR Window. The achromatic Airy disk comparison suggests that beam chromaticity is not the dominant factor, and overall sidelobe level is more important. The theoretical analysis presented in Section \ref{sec:theory} suggests that deeper source subtraction outweighs wider source subtraction, but that both factors are important. In the EoR, at lower redshift, subtracting sources above 1~Jy (intrinsic) out to $\theta_Z=60$ degrees, or 200~mJy out to $\theta_Z=45$ degrees, will be required to access relevant modes of the power spectrum (Figure \ref{fig:angles}).  

In many ways, this work presents the best-case scenario for EoR/CD experiments; no noise, no calibration errors, a complete point source-only sky model, no extended sources, and simplified element patterns to form the station beams. However, it also takes a simplistic view of how one would undertake the EoR/CD experiments, not considering station apodization to control sidelobes, or sub-arrays to gain information on larger scales in the Cosmic Dawn. In the view of this author, station apodization will be required to undertake these experiments effectively, at the expense of sensitivity. Lessons from precursor telescopes such as the MWA, HERA and LOFAR have demonstrated that spectral purity is more important than pure sensitivity. Future work will employ the FEE models to undertake this analysis more rigorously, and help to guide specific details of how the EoR/CD experiments are undertaken.

\subsection* {Acknowledgments}
We would like to thank Randall Wayth for providing the AAVS element patterns and station configuration.
This research was partly supported by the Australian Research Council Centre of Excellence for All Sky Astrophysics in 3 Dimensions (ASTRO 3D), through project number CE170100013. CMT is supported by an ARC Future Fellowship under grant FT180100321.
The International Centre for Radio Astronomy Research (ICRAR) is a Joint Venture of Curtin University and The University of Western Australia, funded by the Western Australian State government. 

\section*{Appendix}
\begin{figure}
\begin{center}
\begin{tabular}{c}
\includegraphics[height=6.5cm]{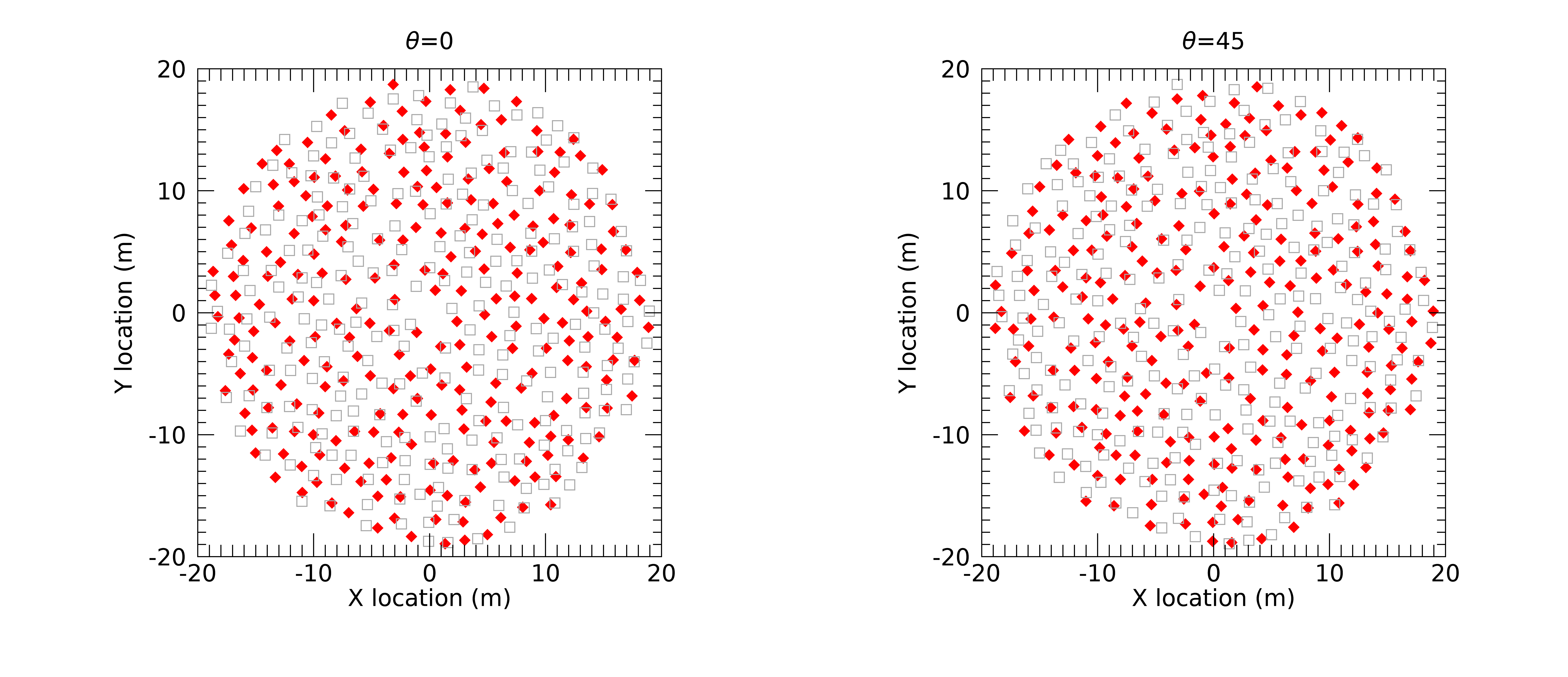}
\end{tabular}
\end{center}
\caption 
{ Two example station layout configurations. (Left) Original AAVS configuration (red diamonds) with grey-scale 45 degree configuration for comparison (squares). (Right) Rotated 45 degree configuration (red diamonds) with grey-scale original configuration for comparison (squares). \label{fig:configs}} 
\end{figure}

Here we display two configurations used in this analysis (chosen from the eight). Figure \ref{fig:configs} (left) displays the actual AAVS configuration (no rotation) in filled red diamonds, overlayed with a grey-scale plot of the 45 degree rotated configuration. The right-hand plot displays the opposite, highlighting the 45 degree configuration, with the actual AAVS layout in grey-scale squares.





\vspace{2ex}\noindent\textbf{Cathryn Trott} is an Associate Professor at Curtin University and the International Centre for Radio Astronomy Research. She received her BSc and PhD degrees in physics from the University of Melbourne in 2001 and 2005, respectively.  She is the author of more than 120 journal papers and has written three book chapters. Her current research interests include Epoch of Reionisation, statistical signal processing, and estimation and detection theory.

\vspace{1ex}

\listoffigures
\listoftables

\end{document}